\documentstyle[preprint,aps]{revtex} %preprint

\input{epsf}
\begin{document}

\let\nopictures=Y

%\draft

%\preprint

\nopagebreak
\title
{
A TWO--EXPONENT MASS-SIZE POWER LAW \\
FOR\\
CELESTIAL OBJECTS
} 
\author{J.\ P\'erez--Mercader\footnote{Also at: IMAFF; CSIC; Serrano, 123; 28006 Madrid.}
\\
%\vspace{.125in}
}  
\address{ 
Laboratorio de Astrof\'{\i}sica Espacial y F\'{\i}sica
Fundamental\\
Apartado 50727\\
28080 Madrid\\
\vspace{.25cm}
{\sl Submitted to Nature}}

\date{{\bf August 19, 1996}}

\maketitle

% \vspace{-6cm}
\vspace{-11.25cm}
\begin{flushright}
LAEFF--96/019\\
August 1996
\end{flushright}
\vspace{7.50cm}
% \vspace{9cm}

\begin{abstract} 

\vspace{-.75cm}

{
The Universe that we know is populated by structures made up of aggregated matter that organizes into a variety of objects; these range from stars to larger objects, such as galaxies or star clusters, composed by stars, gas and dust in gravitational interaction. We show that observations support the existence of a composite (two--exponent) power law relating mass and size for these objects. We briefly discuss these power laws and, in view of the similarity in the values of the exponents, ponder the analogy with power laws in other fields of science such as the Gutenberg--Richter law for earthquakes and the Hutchinson--MacArthur or Damuth laws of ecology. We argue for a potential connection with avalanches, complex systems and punctuated equilibrium, and show that this interpretation of large scale--structure as a self--organized critical system leads to two $predictions$: (a) the large scale structures are fractally distributed and, (b), the fractal dimension is $1.65 \pm 0.25$. Both are borne out by observations.
}

\end{abstract} 

\vspace{.125in}

%\vspace{.5in}

{\sl Subject headings: Gravitation; Cosmology; Complex Phenomena.}

\vspace{.5in}

\narrowtext
% \widetext
% \twocolumn
\nopagebreak

Each point in Figure 1 represents typical {\it (average)} observational data for 
each of the structures known to exist in the Universe whose mass is approximately the mass of the Sun or larger. The abscissa of this 
log--log plot represents the characteristic linear size of the structure 
normalized to the size of the Sun; in the ordinate we have plotted one over the mass of the object normalized to the mass of the Sun, or, what is the same, one over the number of suns that would fit in the structure if all its mass was made up of suns. The actual data 
is presented in Table I.

A very clear pattern of a {\it composite} power law emerges from the figure: 
a power law for ``star--like'' objects {\sc(s)} and a different power law for 
what 
we can generically call\footnote{Because they are aggregates of  ``star--like'' 
objects together with dust, gas, etc..}  ``multistellar'' objects {\sc(ms)}. 
The two
exponents are radically different and they are given by $\tau_{S}= -0.0999$
and by $\tau_{MS}=-2.21$. These values are obtained by a simple chi--squared 
fit to the logarithm of the data of Table I to two different straight lines: one line for the 
{\sc s}--class of objects and another for the {\sc ms}--class of objects. 
The fit properties 
are described in Tables II (for {\sc s}--objects) and III (for {\sc ms}--objects);
we see 
that the fit quality is {\it passable} for {\sc s}--objects and {\it 
excellent}
for {\sc ms}--objects. The fit dependence on $\Omega$ and $h$ is very mild: for $\Omega=0.1$ and $h=1$, $\tau_{MS}=-2.1213$ and $a_{MS}=34.877$, whereas for $\Omega=1.0$ and $h=0.5$, $\tau_{MS}=-2.1836$ and $a_{MS}=36.4538$; with errors very close to what is shown in Table III.

Thus we have the relation 

\begin{displaymath}
\log \left( \frac{M_{\rm object}}{M_{\odot}} \right) = a_{\rm class} +
\tau_{\rm class} \log \left( \frac{l_{\rm object}}{l_{\odot}} \right) 
\end{displaymath}

\noindent
with $M_{\rm object}$ the typical mass of the object, $l_{\rm object}$ its
longitudinal size. The two parameters $a_{\rm class}$ and $\tau_{\rm class}$
have the values quoted in the table for each class. 

The fact that there
are two different power laws for the two classes of objects is a good indication that
objects within each class share a common physical origin which is, in turn,
different for each of the two classes. 

It is a remarkable coincidence that a similar, composite power law, has been found \cite{hutch} to hold for the {\it number of species} of all different kinds of multicellular\footnote{This also extends to bacteria and prokaryotes, as found in Reference \cite{schopf}.} terrestrial animals versus their {\it body length}. Furthermore, the exponents for multicellular animals are very close to what we find in this paper, especially the exponent for the larger--sized species, which is approximately equal \cite{may} to $-2$, and would then correspond to what we have called the {\sc ms}--class of objects.

The existence of the power law for
the {\sc s}--class of objects has been known for many years, and is the celebrated 
mass-radius law\cite{sciamaeddington}. It can be understood using basic features of the
astrophysics of stellar objects and dimensional considerations, as was done
by Eddington who related it to the mass-luminosity ratio. It codifies a
great deal of information on the physics of gravitational collapse and the
nuclear physics of stellar material. This power law holds for about one
order of magnitude in mass and eight orders of magnitude in size. 

The rest of this paper is devoted to the power law for {\sc ms}--objects. This power law is valid from open stellar clusters, the 
simplest type of objects in this class (i.e., the one with the smallest number of components), to the
full Universe, obviously the most complex of the objects in its class, and spans 
twenty orders of magnitude in mass and almost
nine orders of magnitude in linear size as can be seen from Figure 2. The relationship is clearly
non-gaussian (which would show in this log--log plot as a sharp, cut--off curve), and therefore implies the
existence of some scale--free phenomena, {\it perhaps} revealing
the presence of a critical regime\footnote {Since power laws are scale--invariant, they are construed in statistical mechanics 
as the quintessential indicators of criticality\cite{binneyetal}.}, manifesting in the
variety of structures we list under the category of multistellar objects in
Table I. All these objects would then have their origin in a {\it common
mechanism}, possibly related\footnote{An
understanding of this power law can be gained by using the hydrodynamics of many bodies in
gravitational interaction in an expanding Universe. What emerges \cite{jpm} is an
$avalanche$ \cite{avalanche} picture of the large scale structure in the Universe as a particular case, \cite{jpmmgm}, of a very general class of phenomena. See below.} to their
complex many body nature (in the sense that they are made up of {\it many}
star--like objects), which is a property shared by all of them, and
establishes a clear difference between these objects and the objects in the
other class, characterized by being compact\footnote {The term compact is
{\it not} used here in the sense which is used for completely collapsed objects like black holes.} objects. The
presence of ``gaps'' between the structures (we do not know of
any structures intermediate between, e.g., ``globular clusters'' and
``galaxies") points in the direction of
``avalanche'' behavior and, perhaps, some form of ``punctuation'' in the sense of
Elredge and Gould\cite{elredgegould}.  

The ordinate in these plots can also be interpreted, for example, as the $cumulative$ dynamical--mass fraction of suns that would be contained in structures with longitudinal size equal or greater than $l_{object}$. This interpretation of the ordinate {\it together with} the results of our fit, specially for the
{\sc ms}--objects, remind one of the Gutenberg--Richter relation \cite{guttenbergrichter} for earthquakes,
another power law which describes the cumulative number of earthquakes, $N$,
of magnitude\footnote{The magnitude of an earthquake is proportional to the logarithm of the energy released in the quake or contained in the seismic wave amplitude.} greater than or equal to a value $M$ through the expression

\begin {equation}
\log_{10} N=a - b M
\end {equation}

\noindent 
with exponent {\it b} typically slightly greater than 1. This, like punctuated equilibrium, is again a phenomena which can be related to a form of avalanche 
behavior\cite{avalanche}. 

Shifting the ordinate by $\log (M_{\odot}/M_{Universe})$, it represents the  $\log (M_{Universe}/M_{object})$, i.e., the logarithm of the total number of objects of a given dynamical mass fitting in the Universe. Thus one can think of an analogy with Damuth's law \cite{damuth} of ecology which relates the population density, $D$, of moving life--forms\footnote{For comments on the extension of a similar law to plants, see p. 45 of the book by Bonner in Ref. \cite{may}.} on the surface of the Earth to their longitudinal size $L$, according to

\begin {equation}
\log D~\sim ~ \tau_{Damuth} \log L + constant
\end {equation}

\noindent 
with power law exponent $\tau_{Damuth}$. It is very intriguing that as Damuth found, 
$\tau_{Damuth} \approx -2.25$, very close to our $\tau_{MS} =-2.21$ for multistellar objects\footnote {This, of course, was interpreted by Damuth,  using Kleiber's law of 
metabolism, as evidence that there exists a ``pyramid of metabolism".}.

Can one gain quantitative understanding of this power law? Unlike in geophysics or in ecology, where a mathematical framework is less well defined, in cosmology there is a well defined formalism which at least permits one to attack the description of structure formation \cite{padmanabhan}, \cite{colesandsahni}. Within this framework, the equations that describe density perturbations and structure formation in an expanding Universe can be written in a form similar to the Directed Percolation Depinning (DPD) model in 3 + 1 dimensions\cite{jpm}. It is known from computer simulations of DPD--models that they display self--organized critical behavior \cite{bak}, \cite{stanley} (i.e. they lead to power laws) and, in particular, that the probability that an avalanche of time duration $t$ survives scales as

\begin{equation}
P(t) \sim t^{-\tau_{survival}}
\label{survival}
\end{equation}

\noindent
with \cite{stanley} $\tau_{survival} \approx 2.54$, which is reasonably close to the exponent found here for the {\sc ms}--objects. Notice that the (comoving) longitudinal size of the object is related to the time it takes light to cross it by $l=ct$; this is the time entering in the above equation. Furthermore, since the ``number of objects of a given dynamical--mass fitting in the Universe" is the same as the probability that an object of a given dynamical--mass is present in the Universe, we see that our phenomenological {\sc ms}--power law represents Equation (\ref{survival}) with $\tau_{survival} \approx 2.21$. 

If one adopts this ``avalanche" interpretation of the {\sc ms}--power law, there is a $prediction$ that arises from it. As shown by the authors of Ref. \cite{stanley}, in DPD--models the {\it ``shell of unblocked cells in the interface forms a fractal dust"}, which can be related to the distribution of survival times for the avalanches, Equation (\ref{survival}). This fractal dust is packed into ``moving blocks" which behave as quasi--particles and are distributed like a fractal of dimension $d_{dust}$. The dimension of the fractal (or Levy) dust is related to $\tau_{survival}$ through 

\begin{equation}
d_{dust}=z(\tau_{survival}-1) \,\, ,
\end{equation}

\noindent
where the dynamical exponent $z$ is defined as the exponent relating typical time scales to typical length scales, i.e. $t \sim L^z$. In 3+1 dimensions, computer simulations of DPD--models \cite{stanley} give $z=1.36 \pm 0.05$, and therefore the $predicted$ fractal dimension for the dust is $d_{dust}=1.65 \pm 0.25$. The equations of cosmological hydrodynamics are best approximated by the DPD--model for times in the History of the Universe not very much after the decoupling era, and thus one expects that the predictions for scales in the realm of the galaxies and larger of ``avalanche" physics, as described here, will be the most accurate. It is very tantalazing (a) that the ``avalanche" physics interpretation leads to a fractal distribution of the largest structures in the Universe, a fact which has been known since the time of the first large scale surveys and (b) that the value predicted for $d_{dust}$ is so close to what is inferred from observations \cite{peebles} for galaxies, rich clusters and clusters of galaxies, $d_{obs}=1.65 \pm 0.15$.

We end by making several remarks. It is conceivable that the reason why power laws with exponents so similar, but for such diverse and different systems, are present, is because the essential underlying physics is the laws of hydrodynamics together with some form of noise which models the huge number of ``unforeseebles" in a very complex system. The closeness in the values of the exponents, together with the $intuitive$ connection to systems exhibiting Self--Organized Critical behavior, such as earthquakes, the ecology and DPD, together with the correctness of the predictions derived from it leads one to surmise whether the Universe is yet another example of a Self--Organized Critical system, and that this is the general $principle$ behind this phenomenology.

A final question is that one must understand how the two power laws are 
connected together or if they can be understood in terms of a single law 
encompassing both, and why the change of exponent takes place. This requires 
new ways of looking 
at gravitational  phenomena and the large scale structure in the Universe and, like other aspects of large scale structure physics \cite{gang4},
points in the direction of phase transitions and critical phenomena 
where a change of value in the exponent of the power law can take place as a consequence of a phase change.

\section{Acknowledgements}
The author thanks Pere Alberch, Ram Cowsik, Murray Gell--Mann, Terry Goldman, Salman Habib, Dennis Sciama and Geoffrey West for discussion and useful comments.

\newpage

\begin{figure*}
\epsfxsize=15cm
\epsfbox{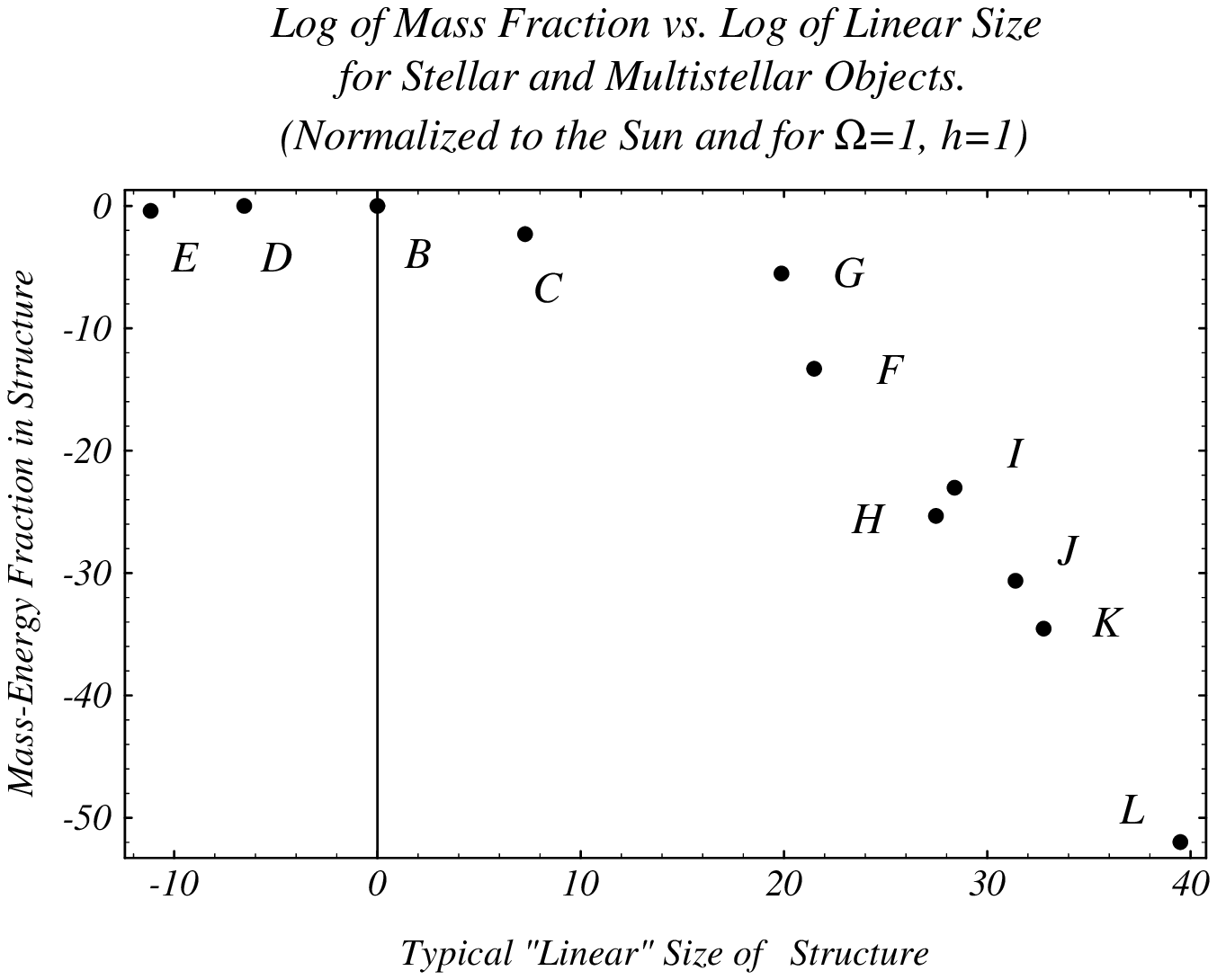}
\caption{Plot of the data in Table I.}
\label{fig:jod1}
\end{figure*}

\begin{figure*}
\epsfxsize=15cm
\epsfbox{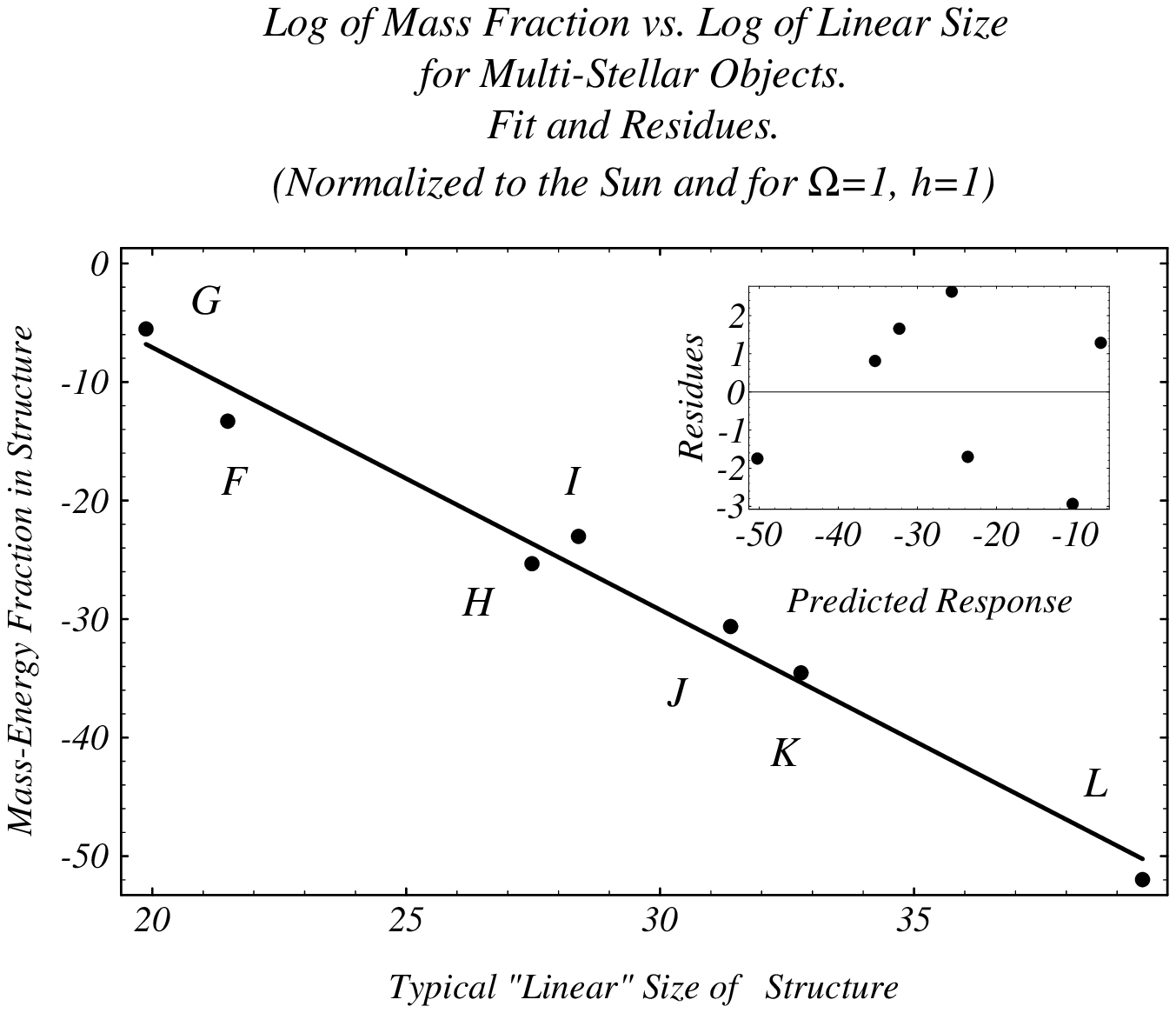}
\caption{Power--law fit and residues for {\sc ms}--objects.}
\label{fig:jod2}
\end{figure*}

\newpage

\begin{table}[t]
\caption{Characteristic longitudinal sizes and $dynamical$ masses for typical celestial objects. These $observational$ data are taken from Reference [1]. To help the reader's intuition we have also included in this table the data on Jupiter. The labels refer to Figure 1. Here $\Omega$ and $h$ are the density parameter for the Universe and the normalized (to $100 \,\,\, km \, sec^{-1} 
\, Mpc^{-1}$) Hubble parameter, respectively.
}
\vspace{0.4cm}
\begin{center}

\begin{tabular}{|c||c|c|c||c|}
{\sc  OBJECT TYPE} & {\sc CLASS} &  {\sc RADIUS} (cm) &  {\sc MASS} (g) &  {\sc LABEL}\\
\hline
 Jupiter & {\sc --} & $6\times 10^9$ & $2\times 10^{30}$ & - \\
\hline
 Neutron Star & {\sc S} & $10^6$ & $3\times 10^{33}$ & E \\
\hline
 White Dwarf & {\sc S} & $10^8$ & $2\times 10^{33}$ & D \\
\hline
 Sun & {\sc S} & $7\times 10^{10}$ & $2\times 10^{33}$ & B \\
\hline
 Red Giant & {\sc S} & $10^{14}$ & $(2-6)\times 10^{34}$ & C \\
\hline
\hline
 Open Cluster & {\sc MC} & $3\times 10^{19}$ & $5\times 10^{35}$ & G \\
\hline
 Globular Cluster & {\sc MC} & $1.5\times 10^{20}$ & $1.2\times 10^{39}$ & F \\
\hline
 Elliptical Galaxy & {\sc MC} & $(1.5-3)\times 10^{23}$ & $2\times (10^{43}-10^{45})$ & I \\
\hline
 Spiral Galaxy & {\sc MC} & $(6-15) \times 10^{22}$ & $2\times (10^{44}-10^{45})$ & H \\
\hline
 Group of Galaxies & {\sc MC} & $3\times 10^{24}$ & $4\times 10^{46}$ & J \\
\hline
 Cluster of Galaxies & {\sc MC} & $1.2\times 10^{25}$ & $2\times 10^{48}$ & K\\
\hline
 Universe & {\sc MC} & $10^{28}/h$ & $7.5\times 10^{55} \Omega/h$ & L \\
\end{tabular}

\end{center}
\end{table}
\newpage

%%%%%%%%%%%%%%%%%%%%%%%%%%%%%%%%%%%%%%%%%%%%

\begin{table}[t]
\caption{Basic Information from the Analysis of Data for Stellar ({\sc s}) objects assuming a Simple Linear Regression equation.}
\vspace{0.4cm}
\begin{center}

(a) Predictive equation
$$
\log \left( \frac{M_{\rm object}}{M_{\odot}} \right) = a_{\rm S} +
\tau_{\rm S} \log \left( \frac{l_{\rm object}}{l_{\odot}} \right)
$$

\vspace{0.4cm}
(b)Estimates of Regression Coefficients and their standard errors in power law for Stellar objects.
\vspace{0.4cm}

\begin{tabular}{|l||c|c|}
Parameter& $a_{\rm S}$ & $\tau_{\rm S}$ \\
\hline
Parameter estimate& $-0.9379$ & $-0.0999$ \\
\hline
Standard error of & & \\
parameter estimate & $0.493505$  & $0.0665$\\
\end{tabular}

\vspace{0.4cm}
(c) Analysis of variance.
\vspace{0.4cm}

\begin{tabular}{|l||c|c|c|c|c|}
Source of variation& D.F & S.S. & M.S. & $F$ & $p$\\
\hline
Regression& $1$ & $1.9258$ & $1.9258$ & $2.2563$ & $0.2719$ $>$ $0.01$ \\
Residual& $2$ & $1.7071$ & $0.8535$ & &  \\
\hline
Total &$3$ & $3.6329$ &  & &   \\
\end{tabular}

\vspace{0.4cm}
(c) Percentage of variation explained by explanatory variable.
\vspace{0.4cm}
$$
R^2=53 \%
$$
$$
R^2_{\rm adj}=29.5 \%
$$

\end{center}
\end{table}
%%%%%%%%%%%%%%%%%%%%%%%%%%%%%%%%%%%%%%%%%%%%

\begin{table}[t]
\caption{Basic Information from the Analysis of Data for Multistellar ({\sc ms}) Objects assuming a Simple Linear Regression equation and that $\Omega =1$ and $h=1$. See also Figure 2.}
\vspace{0.4cm}
\begin{center}

(a) Predictive equation
$$
\log \left( \frac{M_{\rm object}}{M_{\odot}} \right) = a_{\rm MS} +
\tau_{\rm MS} \log \left( \frac{l_{\rm object}}{l_{\odot}} \right)
$$

\vspace{0.4cm}
(b)Estimates of Regression Coefficients and their standard errors in power law for Multistellar objects.
\vspace{0.4cm}

\begin{tabular}{|l||c|c|}
Parameter& $a_{\rm MS}$ & $\tau_{\rm MS}$ \\
\hline
Parameter estimate& $35.1723$ & $-2.21274$ \\
\hline
Standard error of & & \\
parameter estimate & $4.10181$  & $0.139667$\\
\end{tabular}

\vspace{0.4cm}
(c) Analysis of variance.
\vspace{0.4cm}

\begin{tabular}{|l||c|c|c|c|c|}
Source of variation& D.F & S.S. & M.S. & $F$ & $p$\\
\hline
Regression& $1$ & $1331.74$ & $1331.74$ & $251.01$ & $2 \times 10^{-5} < <$ $0.01$ \\
Residual& $5$ & $26.5287$ & $5.30574$ & &  \\
\hline
Total & $6$ & $1358.27$ &  & &   \\
\end{tabular}

\vspace{0.4cm}
(c) Percentage of variation explained by explanatory variable.
\vspace{0.4cm}
$$
R^2=98.0 \%
$$
$$
R^2_{\rm adj}=97.7 \%
$$

\end{center}
\end{table}

\end {document}